\documentclass{article}
\usepackage[utf8]{inputenc}
\usepackage{graphicx}
\usepackage{float}
 \usepackage{url}
 \usepackage{xcolor}
 \usepackage{amsmath}
 \usepackage{comment}
 \usepackage{amsmath}

\title{A Computational Model for Measuring Adaptability Among U.S. Farmers: Evidence from 1997–2022}

\author{
\texttt{Hossein Sabzian}\\
\texttt{hossein.sabzian@maine.edu}\\
\texttt{The University of Maine, Orono 04473, ME, USA}
}

\date{\date{}}

\begin{document}

\maketitle

\section{Abstract}

Agricultural crops are a type of cultural trait and the way farmers of US counties select them can itself result in county-level cultural traits. Using real-world data from 1997 to 2022, we have developed a systematic framework to study the selective mechanisms behind these traits. Our findings indicate that environmental payoff-biased selection has driven counties to adopt traits that maximize their adaptability and yield within their specific environments. These empirical results align with existing theoretical literature \cite{bentley2004random, henrich2001cultural}. Additionally, a clear long-term selective trend is evident, showing that US counties are gradually developing a specific set of more complex combinatorial traits, which provide greater payoffs by enhancing the farmers' environmental adaptability. This study serves as a strong case for empirically modeling the cultural evolutionary processes among US farmers.

\section{Introduction}

Culture is one of the most complex collective phenomena we know. The capacity of our species to produce and accumulate culture (i.e. the collection of socially transmitted attitudes, beliefs, knowledge, skills, technologies, values, etc.) has allowed us to inhabit virtually every part of the planet, making culture a corner- stone to the ecological success of our species\cite{henrich2016secret, van2016primate} . Moreover, no other species produce as many different and complex cultural products and is undoubtedly as reliant on them as we are \cite{boyd1996culture, van2016primate}. As such, the study of culture and how it changes has been central to the research programme of many disciplines, from the humanities and social sciences to the natural sciences.

For over a century, evolutionary theory has been applied to the study of culture. In the early days of cultural evolution research, particularly before the 1970s, the approaches were largely narrative-based and interpretive. These traditional methods did not offer quantitative predictions about how cultural traits (such as behaviors, ideas, and artifacts) would vary and be distributed within a population. Modern cultural evolution research diverges significantly from these earlier approaches by adopting a quantitative and mathematical framework.

The foundations for this shift were laid by early pioneers like Gulick (1905)\cite{gulick1905evolution} and Binford (1963)\cite{binford1963red} . However, it was not until 1981 that a comprehensive quantitative framework was developed by Cavalli-Sforza and Feldman\cite{cavalli1981cultural}. These evolutionary scientists borrowed mathematical tools from population genetics to predict how the social transmission of information influences cultural dynamics at the population level. Their work introduced two key concepts: 
\begin{enumerate}
    \item Cultural selection, analogous to natural selection, which describes the differential reproductive success of cultural traits and is considered deterministic.
    \item Cultural drift, analogous to genetic drift, which accounts for the role of chance in cultural change.
\end{enumerate}

Following their seminal work, the field saw a surge of studies in anthropology and archaeology, as reviewed by Mesoudi (2011)\cite{mesoudi2011cultural}and Lewens (2015)\cite{lewens2015cultural}. These studies primarily focused on social transmission mechanisms, such as the relative importance of vertical transmission (from parents to offspring) compared to horizontal transmission (from peers). Building on Cavalli-Sforza and Feldman's framework, Boyd and Richerson (1985)\cite{boyd1988culture}  and their students introduced various modes of cultural selection, termed biased cultural transmission. These include prestige-bias, conformist-bias, and payoff-biased social transmission\cite{henrich2003evolution, henrich2001cultural}. 

The concept of cultural selection has sparked significant interest and debate within the human sciences, particularly between the schools of cultural anthropology\cite{acerbi2015if} and cognitive anthropology\cite{claidiere2014darwinian}.
In contrast to the widespread acceptance of cultural selection, the concept of cultural drift\_or unbiased selection has not gained as much traction and remains underexplored\cite{billiard2018stochasticity}. Neutral models, which are a subset of stochastic models where population change is not directed towards any particular outcome, have seen productive use in archaeology and anthropology. However, there has been little debate over the relative roles of selection and drift in cultural evolution, unlike in biological sciences where this has been a long-standing issue. In population genetics and ecology, the stochastic revolution has shifted the default assumption from selection to stochastic processes such as genetic drift and demographic stochasticity\cite{kimura1979neutral, hubbell2011unified}. Selection models now must demonstrate greater explanatory power than neutral models to account for observed data.

In this paper, we model a cultural evolutionary process within the US agricultural system using real-world data. We aim to illuminate the possible mechanisms of cultural selection (whether it is biased or unbiased in our case) and identify its underlying sources. The paper is structured as follows:
\begin{enumerate}
    \item Introduction to the cultural systems studied: An overview of the cultural traits and systems examined.
    \item Selection mechanism in cultural context: A discussion of selection mechanisms within cultural evolution.
    \item Construction of trait space from real-world data: The creation and analysis of trait space using empirical data.
    \item Testing the memorylessness property of trait space: An examination of whether trait space exhibits the Markov property.
    \item Analysis of trait diffusion over time: Exploration of equilibrium, selection source and adaptive evolution of traits over time.
    \item A final conclusion: Discussion of findings, possible implications and future directions.
\end{enumerate}

\textbf{\textit{This structured approach allows us to comprehensively investigate the dynamics of cultural evolution and the factors driving these processes.}}

\section{Cultural Selection}

Selection in the cultural case refers to learning forces that favor some behavioral variants over others. For example, people probably prefer to imitate the successful and this favors behaviors that lead to success \cite{dunbar2007oxford}. Cultural selection is defined as any situation in which one trait is more likely to be acquired and passed on than an alternative cultural trait \cite{mesoudi2011cultural}. There are various mechanisms of cultural selection, including model-biased, content-biased, and frequency-biased selection. Model-biased selection involves preferentially adopting traits based on the characteristics of the model, such as age, seniority, or similarity. Content-biased selection entails preferentially adopting traits based on their intrinsic attractiveness. Frequency-biased selection involves preferentially adopting traits based on their prevalence, such as through conformity, which is the practice of copying the most popular trait \cite{kendal2018social}.

Most studies on cultural selection are focused on how people try to choose among cultural variants \cite{mesoudi2011cultural, fog1999cultural, henrich2001cultural}, and a study about whether current trait(s) that people have adopted so far can affect their future selection is still lacking. In choosing alternative traits over each other, The probability that we choose the next cultural trait can depend on which cultural trait we are currently bearing. Those who move from s county-style music to hip-hop music style may be following different mechanisms than those who are moving from Rock style style to hip-hop style. 

Therefore, for a trait space \( S=\{trait_1, trait_2, trait_3 ,trait_x\ldots\} \) where all traits are somewhat alternative to each other, The following probability has to be well test.

\begin{equation}
P(X_{t+1} = trait_{x_{t+1}} \mid X_t = trait_{x_{t}}, X_{t-1} = trait_{x_{t-1}}, \ldots, X_0 =  trait_{x_{0}}) = r  
\label{eq:realtion}
\end{equation}

The relation \ref{eq:realtion} encodes the probability of selecting $trait_{x_{t+1}}$ given the chain of selected traits including currently adopted trait $trait_{x_{t}}$, and previously selected ones such as $trait_{x_{t-1}}$ ,$trait_{x_{t-2}}$ , $trait_{x_{0}}$.

For $r$, two relations are possible. On one side, if we have relation \ref{eq:realtion1} for $r$, it means the selection of $trait_{x_{t+1}}$ is dependent on the entire chain of traits that have been selected so far by person (no-Markovian).
\begin{equation}
r = P(X_{t+1} = trait_{x_{t+1}} \mid X_t = trait_{x_{t}}, X_{t-1} = trait_{x_{t-1}}, \ldots, X_0 =  trait_{x_{0}})
\label{eq:realtion1}
\end{equation}

On the opposite side, if we have the relation \ref{eq:realtion2} for $r$, it  shows the selection of $trait_{x_{t+1}}$ just depends on currently selected trait $trait_{x_{t}}$ and the preceding chain's elements has no importance (Markovian), so this process is weakly memoryless in a way that the future possible trait is only dependent on the current trait, not its previous ones despite the fact that each cultural trait per se is a learnable pack of information.

\begin{equation}
 r =  P(X_{t+1} = y \mid X_t = x)
 \label{eq:realtion2}
\end{equation}

 \textbf{\textit{ Technically, the purpose of this study is to test relation \ref{eq:realtion} and discuss its implications.}}

\section{Cultural Trait}

Cultural evolution is an evolutionary theory of social change. It follows from the definition of culture as "\textit{information capable of affecting individuals' behavior that they acquire from other members of their species through teaching, imitation and other forms of social transmission}” \cite{richerson2008not}. As the constituent of culture, a cultural trait is a piece of information or behavior that is socially transmitted within a population \cite{mesoudi2011cultural, mesoudi2018cumulative, mesoudi2021cultural}. Multiple selective mechanisms and transmission pathways exist for choosing, learning, modifying, and transmitting these traits within and between populations\cite{smolla2021underappreciated}.

Each year, US farmers plant a variety of crops (52 different crops). These crops are distributed among US counties (3143 US counties) based on the cultural practices of the farmers. The reproduction and distribution of crops are not solely determined by their inherent biological properties interacting with the environment; rather, they are significantly influenced by cultural factors and selective processes. The outputs (harvested value in bushels) are also determined by cultural practices such as irrigation and fertilizer use.

Agriculture, as a system\cite{altman2019understanding} is a crystal clear example of cumulative cultural evolution. Over successive generations, humans have refined and modified agricultural practices, leading to increasingly complex and efficient systems. This process has involved innovations in tools, techniques, and social organization, all of which have been transmitted and modified across generations.

In this context, the way farmers in each county select and plant crops forms a combinatorial trait that can be studied as a type of cultural trait. These traits are developed to enhance farmers' adaptability and increase their output. From our data (1997 to 2022), 26 cultural traits, labeled trait\_1 to trait\_26, have been extracted, where trait\_n indicates the combination of n crops planted in each county each year. Counties can modify or improve these traits over time to adapt actively and effectively.

\begin{table}[H]
\centering
\includegraphics[width=10cm]{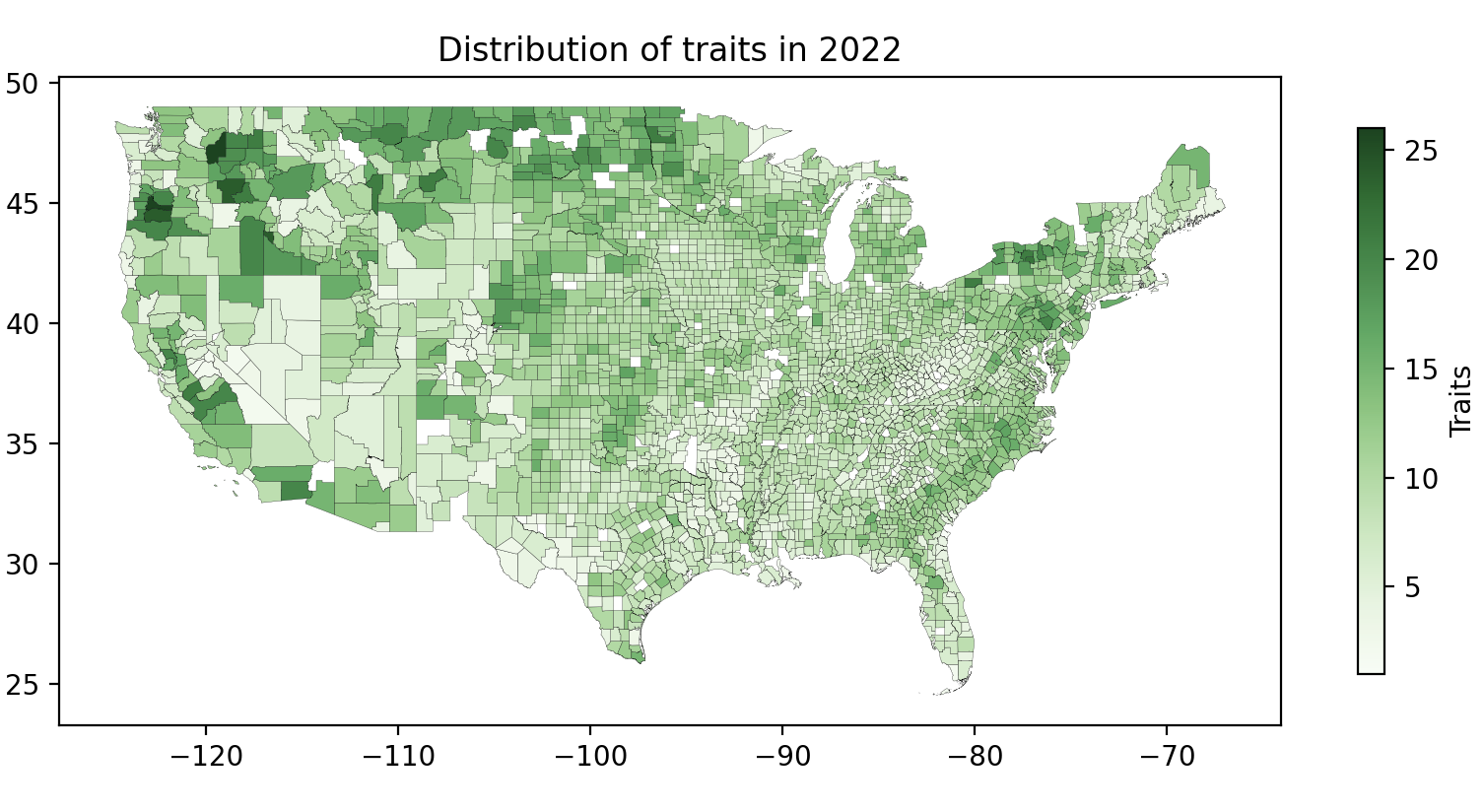}
\caption{Traits in 2022}
\label{tab:traitdis}
\end{table}

\textbf{\textit{This study aims to examine the selective structure among US counties to understand how farmers in each county select crops and adaptively evolve over time.}}

\section{Trait Space}

Using USDA census data, A trait space is made from 15715 recorded data (all US counties from 2002, 2007, 2012, 2017, 2022). Then, this trait space has been transformed into a transition matrix ($M$) which quantifies the switching probabilities among traits (see table \ref{tab:trans}). Understanding the possibility of switching between traits reveals fundamental insight on how likely a trait is to be copied and imitated by other people. 

\begin{table}[H]
\centering
\includegraphics[width=12cm, height=3cm]{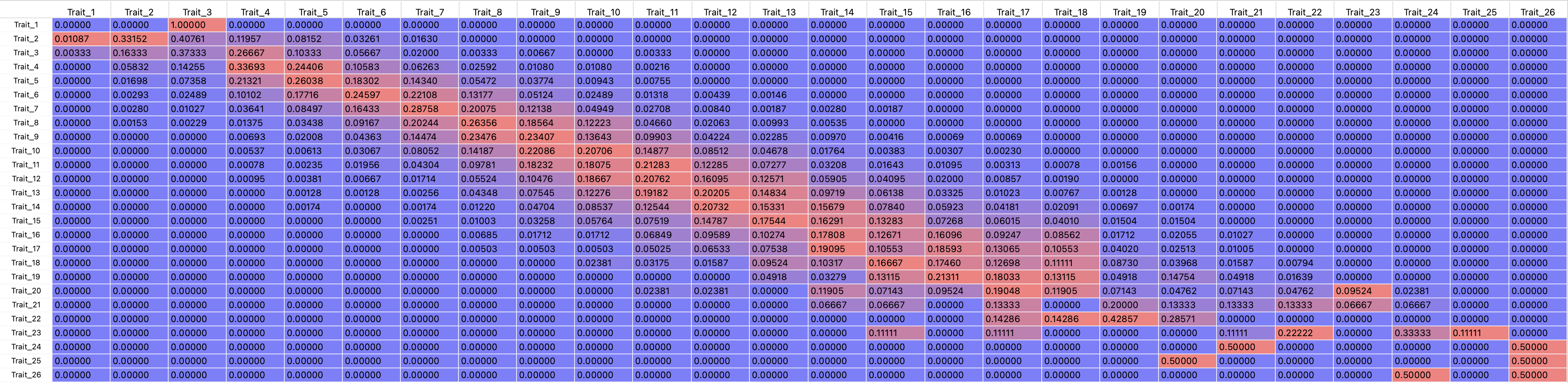}
\caption{Transition matrix of traits}
\label{tab:trans}
\end{table}

\begin{figure}[H]
\centering
\includegraphics[width=12cm]{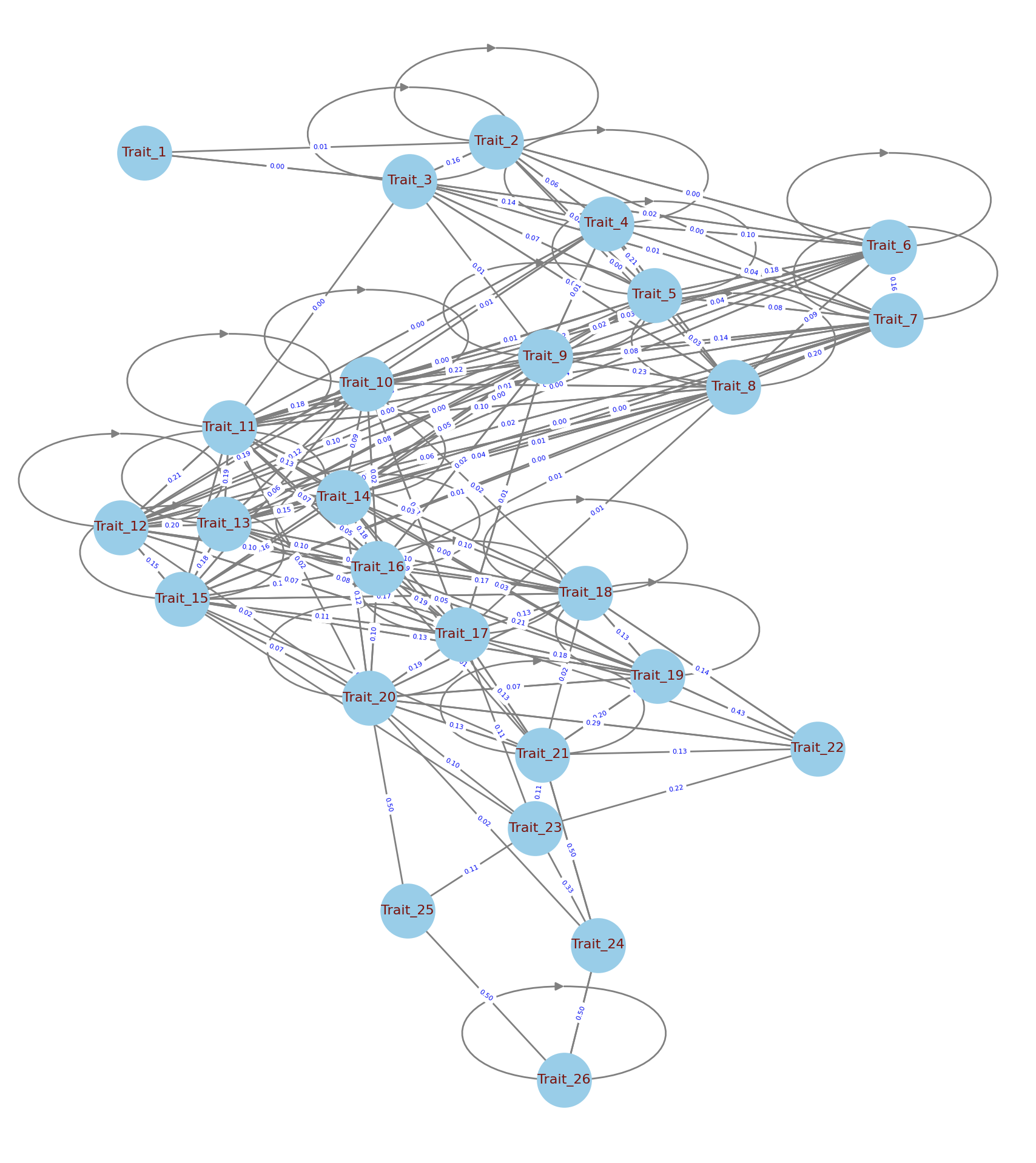}
\caption{Graph of traits}
\label{fig:tran}
\end{figure}

 Each element of the matrix represents the probability of moving from one state to another in a single time step. Key characteristics of a transition matrix are as following:
\begin{itemize}
    \item Square Matrix: The matrix is square, meaning it has the same number of rows and columns. Each row and column corresponds to a state in the Markov chain.
    \item Non-negative Entries: All entries in the transition matrix are non-negative and represent probabilities.
    \item Rows Sum to One: Each row in the transition matrix sums to one, reflecting that the total probability of transitioning from a given state to all possible next states must equal one.
\end{itemize}
The transition matrix is represented as a chain in figure \ref{fig:tran} and those traits that have self-reference arrows, have the probability of transitioning to themselves.

\section{Testing memorylessness}

To test the weak memorylessness of transition matrix, the Chapman-Kolmogorov equation\footnote{Its differential form is a master equation used for modeling continuous-time evolutionary systems.} is used \cite{bowman2013introduction}. This equation is fundamental in verifying the memorylessness of state chains. For a chain with state space \( S \) and transition probabilities \( P_{ij}(t) \), the Chapman-Kolmogorov equation says that for any two time steps \( t \) and \( s \) and for any states \( i \) and \( j \) in the state space we have:

\begin{equation}
P_{ij}(t+s) = \sum_{k \in S} P_{ik}(t) P_{kj}(s)
 \label{eq:champ}
\end{equation}

where:
\begin{itemize}
    \item \( i \): The starting state.
    \item \( j \): The destination state.
    \item \( t \): The number of steps from state \( i \) to state \( k \).
    \item \( s \): The number of steps from state \( k \) to state \( j \).
    \item \( k \): The intermediate state through which the process might pass. It is summed over all possible states \( k \) in the state space \( S \).
    \item \( P_{ij}(t+s) \) is the probability of transitioning from state \( i \) to state \( j \) in \( t+s \) steps.
    \item \( P_{ik}(t) \) is the probability of transitioning from state \( i \) to state \( k \) in \( t \) steps.
    \item \( P_{kj}(s) \) is the probability of transitioning from state \( k \) to state \( j \) in \( s \) steps.
\end{itemize}
 The summation is over all possible intermediate states \( k \). The intermediate state \( k \) essentially allows us to break down the multi-step transition probability into smaller, more manageable probabilities as following:
 \begin{itemize}
     \item First, transition from state \( i \) to state \( k \) in \( t \) steps.
     \item Then, transition from state \( k \) to state \( j \) in \( s \) steps.
 \end{itemize}
By summing over all possible intermediate states \( k \), we account for all possible paths that the process can take from \( i \) to \( j \) in \( t+s \) steps.

After computing equation \ref{eq:champ} for 2-step and 10-step transition matrices and  comparing them with directly computed transition matrices for 2-step ($M^{2}$) and 10-step ($M^{10}$), it is shown that the transition matrix ($M$) has weak memoryless property. Therefore the relation \ref{eq:realtion2} is confirmed which shows the selection of a future possible trait $trait_{x_{t+1}}$, just depends on currently selected trait $trait_{x_{t}}$ and not on the entire history of states (previously selected traits) that led to it. \textbf{\textit{This means that the probability of transitioning to a new state is solely determined by the current state, regardless of what happened before. Therefore, what cultural farming trait that a US farmer is going to select for is dependent on a trait he or she is currently bearing.}}

\begin{figure}[H]
\centering
\includegraphics[width=12cm]{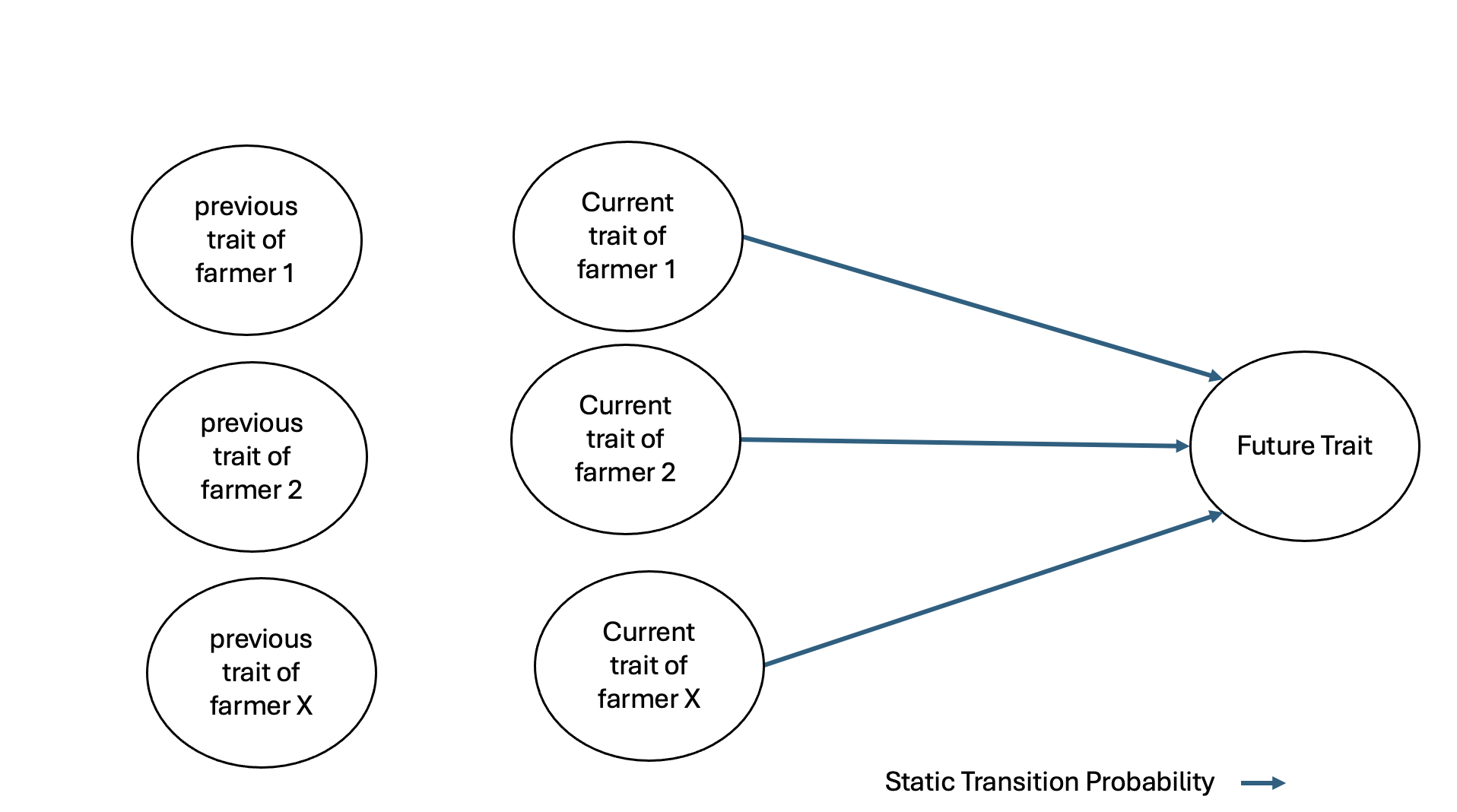}
\caption{Selection based on current situation}
\label{fig:transprob}
\end{figure}

It can be simply depicted in the figure \ref{fig:transprob} which shows the selection of a future possible trait (for farmers) is only dependent on the present situation (trait) they have and all transition probabilities are static.

\section{Diffusion over time}

Using equation \ref{eq:diffe},\footnote{A first-order homogeneous difference equation:} the chain is run for 60 time-steps (60 years) for all crops in US agricultural system. For making its initial state, The 1997 data is used as the base year (\textit{See appendix B}).

\begin{equation}
S(t+1) = M*S(t)
 \label{eq:diffe}
\end{equation}

The table \ref{tab:distristate} shows the distribution matrix of each trait for both first 10 time-steps (from $timestep_{0}$ to $timestep_{10}$) of system and last 10 time-steps of system (from $timestep_{49}$ to $timestep_{59}$). The time-step $0$ is extracted from baseline year (1997). The matrix is sorted based on difference factor (diff) which is the gap between stationary state and initial state $timestep_{0}$.

\begin{table}[H]
  \centering
  \includegraphics[width=12cm, height=4cm]{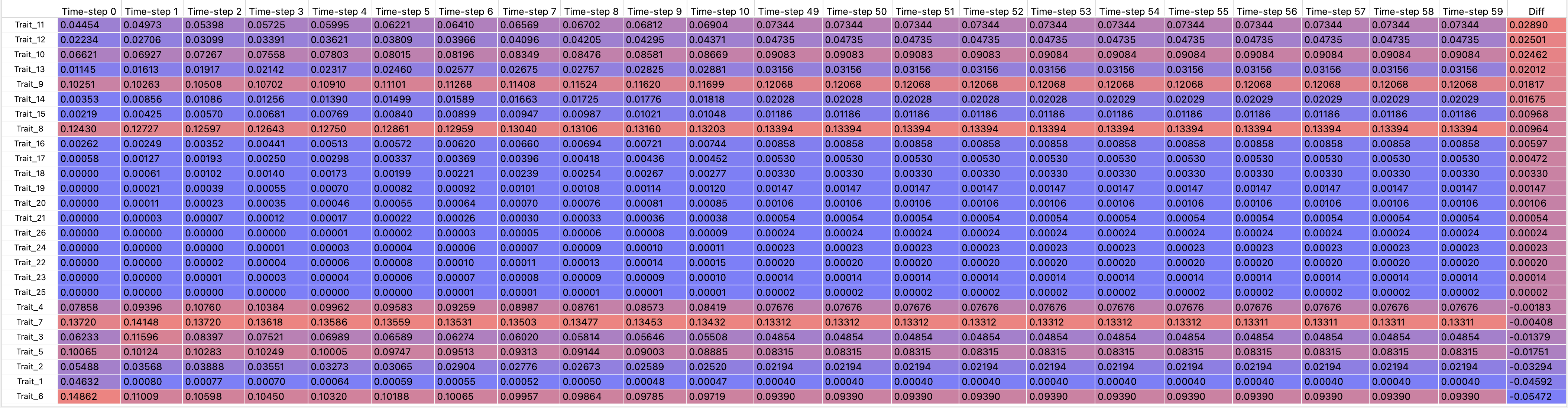}
  \caption{Distribution of traits over long time steps}
  \label{tab:distristate}
\end{table}

\subsection{Equilibrium and selection }

\begin{figure}[H]
  \centering
  \includegraphics[width=12cm]{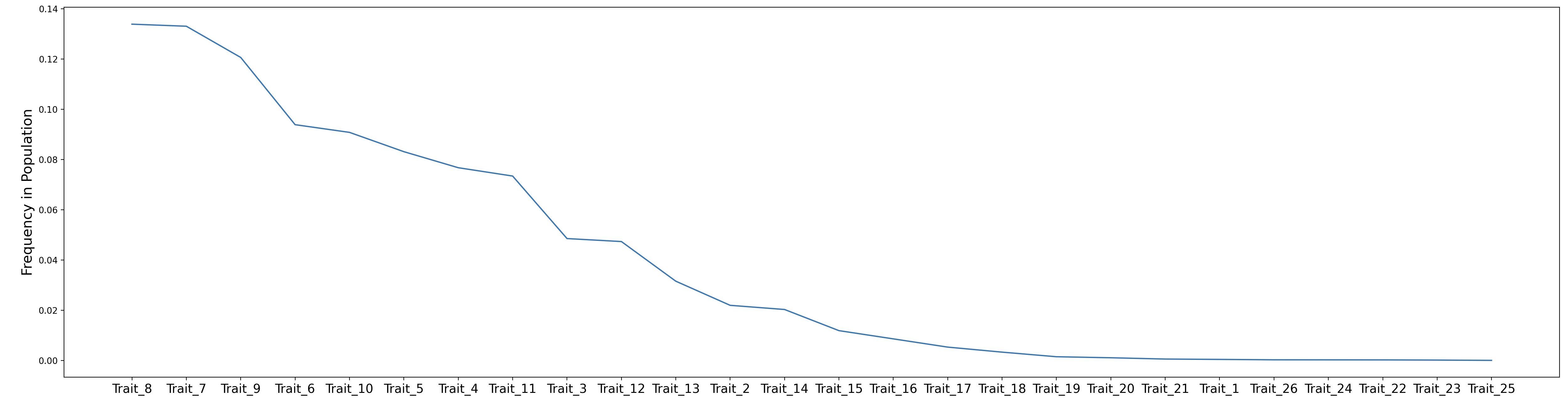}
  \caption{traits selection probability after long time}
  \label{fig:traitdensity}
\end{figure}

The stationary state (SS) of system is in $timestep_{56}$ which shows the equilibrium state of US agricultural system over time, For example, trait 8 has highest score in stationary state meaning that in the long run, US agricultural system will be bearing trait\_8 for 0.13394 of time while it is almost 0.00002 of time for trait\_25. \textbf{\textit{Therefore, those traits that have higher positive scores in the stationary state are going to be more copied/adopted in the future(even if there are maladaptive})}  The probability of selection for each traits in long run is shown as figure \ref{fig:traitdensity} with trait\_8 having highest score and trait\_25 has the lowest one.

The Figure \ref{fig:traitdensity} forms a long-tailed (power-like) structure and according to literature \cite{bentley2004random, mesoudi2021cultural}, unbiased selection can result in power-like distribution for traits adoption over successive generations and that is strongly shown in our model.\textbf{ \textit{So, the way farmers select future traits includes an amount of randomness.}} 

Although the selection of cultural traits by farmers may appear random and ultimately result in a long-tailed distribution of traits, understanding the underlying mechanisms and the nature of this "randomness" is crucial. Therefore, the following section explores the potential sources of this seemingly random selection.

\subsection{Source of Selection }

\begin{figure}[H]
  \centering
  \includegraphics[width=10cm, height=10cm]{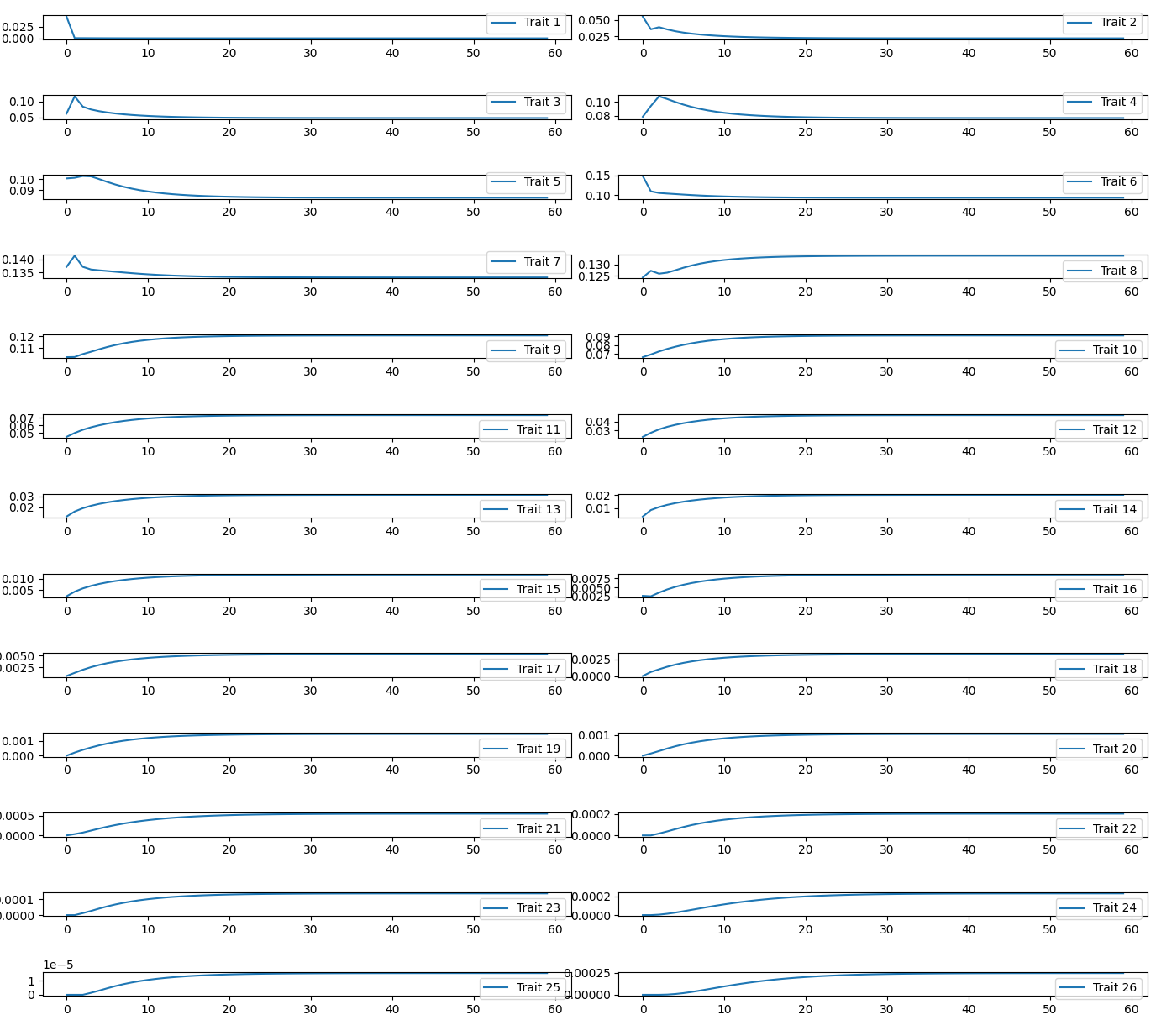}
  \caption{Traits diffusion over time }
  \label{fig:alldif}
\end{figure}

\begin{figure}[H]
  \centering
  \includegraphics[width=12cm]{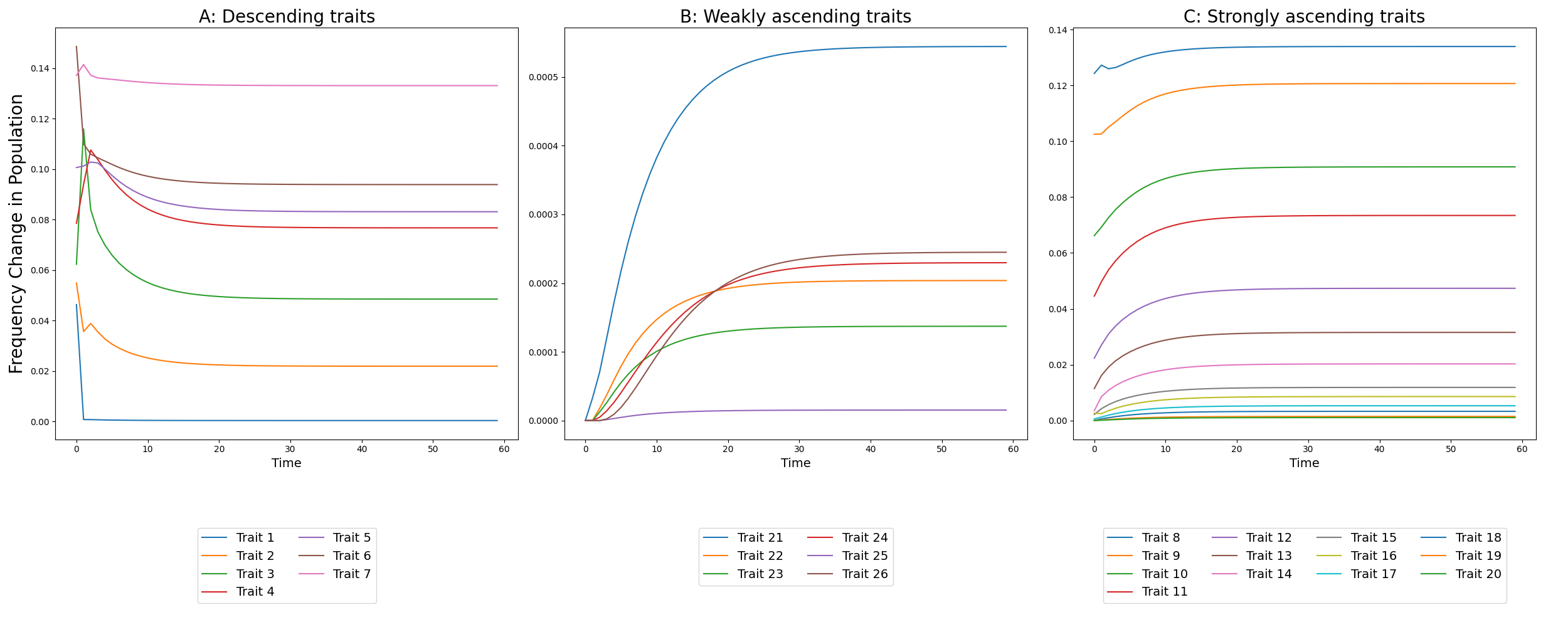}
  \caption{Groups of traits diffusion pattern. A: Declining traits which are those traits that are losing momentum (all traits in maladaptive region of 
 figure \ref{fig:traitsuccess}), B: Weakly ascending traits which are those traits that are gaining momentum with a slow rate(some of those traits in adaptive region of  figure \ref{fig:traitsuccess}), and C: Strongly ascending ascending which are traits that are losing momentum  with a fast rate(some of those traits in adaptive region of figure \ref{fig:traitsuccess}) }
  \label{fig:marko}
\end{figure}

Cultural adaptation occurs through various mechanisms: inadvertent differential survival of populations with advantageous practices, conscious responses to perceived hazards, and random chance \cite{henrich2001cultural}. Understanding why selection process is random or includes randomness can be explained by the structure of the trait space (S) from which the initial transition matrix (M)(see table \ref{tab:trans}) was extracted. The trait space (S) consists of trait sequences over five years (2002, 2007, 2012, 2017, 2022). Since the transition matrix (M) derived from the trait space (S) is Markovian , the force behind its arrangement (i.e., sequences in trait space (S)) is likely stochastic, indicating a random selection process. The diffusion patterns of traits can help us understand the possible force behind random selection process\cite{henrich2001cultural, mesoudi2021cultural}.Based on difference factor of traits, The diffusion of all traits is depicted in figure \ref{fig:alldif}. All diffusion patterns are grouped in \ref{fig:marko} including three groups of of declining traits, weakly ascending and strongly ascending.

As observed, declining traits exhibit an asymptotically decreasing pattern, forming concave curves. In contrast, all ascending traits (both weak and strong traits that are 0.73 of all traits) demonstrate an increasing momentum, displaying an R-curve pattern. According to literature\cite{henrich2001cultural}, R-curve traits are more likely to result from \textbf{\textit{environmental learning model or environmental payoff-biased selection}}. Therefore, the source of random selection is likely to be environmental or related to environmental payoff bias, consequently this force is responsible for shaping the arrangement of trait space (S) elements over the empirically available years (2002 to 2022).

\subsection{Environmental payoff-biased selection and direction of adaptation}

Environment-biased payoff is the strongest possible source for selection of cultural traits in our study. It means that farmers select for those traits that can increase their payoff under environment they are in. So fitting with environment is a major concern for farmers and by changing the combination of crops over years, counties actually trying to get well fit with environment and increase the payoff of planted crop (i.e, they want to harvest more).

Environment-biased payoff selection has led counties to choose traits that maximize their adaptability and yield within their specific environments. The declining trend observed for certain crops (e.g., 1, 2, 3, 4, 5, 6, 7) indicates that the overall system is shifting towards higher-order traits, particularly those that show a strong upward trend (as illustrated in Figure \ref{fig:marko}). These traits are preferred because they provide greater payoffs by enhancing the farmers' environmental fitness.

\subsection{Adaptive Evolution}

For calling something adaptive\_or maladaptive, a number of measures have been used such as survival, relative abundance, rate of population increase, adoption, numerical stability and geographical range but these measures cannot be complete when context and time is not considered\cite{ellen20018}. Being adaptive\_or maladaptive is context-specific, one trait can be adaptive in one context while maladaptive in another context\cite{el2014cultural}. Since US is a big country including more than 3000 counties, No single cultural farming trait is adaptive for all of them that is why we see the presence of multiple traits for US farmers in each time step. The adaptation is also time scale-dependent meaning that The adaptation in a short time differs from long time in a case that sometimes short-time adaptations may turn out to be maldaptive in the long time, for instance using some fertilizers for increasing crops yields may be an adaptive trait for the short time but it will cause soil erosion in the long time and can be a maldaoptive trait.  It is also the case for farmers cultural traits if we pick adoption as the measure of adaptation, In one year, a typical trait may be widely adopted but it will gradually be rejected by some farmers in a long run, for instance, see trait\_6 in table \ref{tab:distristate}, it has the adoption value of 0.14862 in the initial time timestep\(t_0\) (year 1997) that has declined to 0.09390 in timestep\(t_{59}\). This point technically implies the idea of fitness over time that can be captured in absolute form of difference factor.In table \ref{tab:distristate}, the difference factor (Diff) represents the gap between the stationary state and the initial state at timestep\(t_0\). A positive value indicates that a trait has shown increasing adoption over time, while a negative value indicates a decline.  For example, trait\_11 has a higher adoption rate (0.07) compared to its initial state value (0.04). Adaptive region includes all those traits that have gianed a positive adoption over time while the maladaptive traits are the opposite. 

\begin{figure}[H]
  \centering
  \includegraphics[width=12cm]{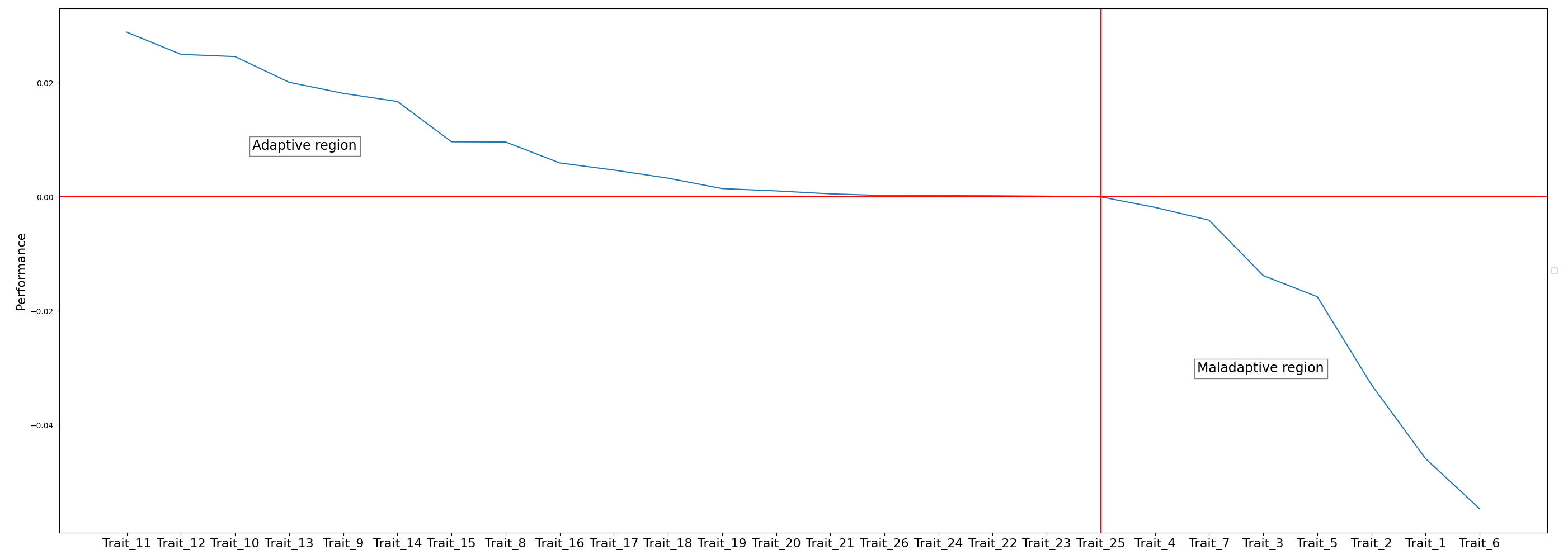}
  \caption{Adaptive and maladaptive traits after a long time}
  \label{fig:traitsuccess}
\end{figure}

This difference factor (Diff) signifies population-level adaptation over time. Figure \ref{fig:traitsuccess} illustrates performance of all traits and those traits that have a positive difference factor are adaptive and those who do not have are maladaptive traits. 
Therefore, those counties that are having these traits in the lung run are maladaptive.
This point shows that even in the long run, the system can be in a maladaptive state, for example, farmers have trait\_4 for 0.07676 of time, 0.13311 of time for trait\_7, 0.04854 of time for trait\_3, 0.08315 of time for trait\_5,0.02194 of time for trait\_2, 00.00040 of time for trait\_1 and 0.09390 of time for trait\_6. \textbf{\textit{This indicates that about 0.4474 of time, US agriculturtal system will be maldaptive. In other words, about 0.4474 of US farmers will be bearing maladaptive traits in the lung run.}}

\section{Conclusion}

Using real-world data from US agriculture (1997 to 2022 at 5-year intervals), we derived 26 cultural traits that farmers in various counties have developed to increase their adaptability over time. These traits indicate the diversity of crops planted annually, and counties continuously strive to maintain or regulate this diversity to achieve the highest adaptability. Transitioning from one trait to another is probabilistic (non-linear) and largely depends on the traits currently adopted by the counties. Therefore, in selecting future cultural traits, counties exhibit a form of weak memorylessness, heavily relying on their current state when deciding on future cultural traits. The selective process appears to be somewhat random, with environment-biased payoff being the primary driver of selection. These empirical findings align with previous literature \cite{bentley2004random, henrich2001cultural}. Our results show that environment-biased payoff selection has led counties to choose traits that maximize their adaptability and yield within their specific environments.

A clear selective trend is observable among US counties over the long term. The declining trend for certain traits (e.g., trait\_1, trait\_2, trait\_3, trait\_4, trait\_5, trait\_6, trait\_7) indicates that the overall system is shifting towards higher-order traits, particularly those showing a strong upward trend (as illustrated in Figure \ref{fig:marko}). These traits are preferred because they provide greater payoffs by enhancing the farmers' environmental fitness. Consequently, counties are gradually developing more complex combinatorial traits over time, which are increasingly centered around trait\_8, trait\_9, trait\_10, and trait\_11 (this trend can be observed in empirical data by comparing trait distributions from 1997 to 2022; see Appendix A).

Finally, although the system exhibits a weak Markov property in this case study, this does not preclude the possibility of transmission and learning within the system over time. For transmission, there may be patterns of trait transmission, for example, where nearby counties exhibit common traits over time. In terms of learning, unlike the current static transition probabilities, these can be dynamically updated based on information that counties learn over time.Additionally, although the environment is shown to be the major force behind selection, there is still a need to: I) find more evidence for it, and II) understand other possible selective processes that have not been thoroughly explored.

\subsection*{Appendix A : Underlying Distribution of Data}

 For density estimation, kernel density estimation (KDE) has been used because we have no assumption about underlying distribution of data unlike maximum likelihood method (MLM) which has assumption of underlying parametric model and Requires a specified parametric form of the distribution.
 
\begin{figure}[H]
  \centering
  \includegraphics[width=12cm]{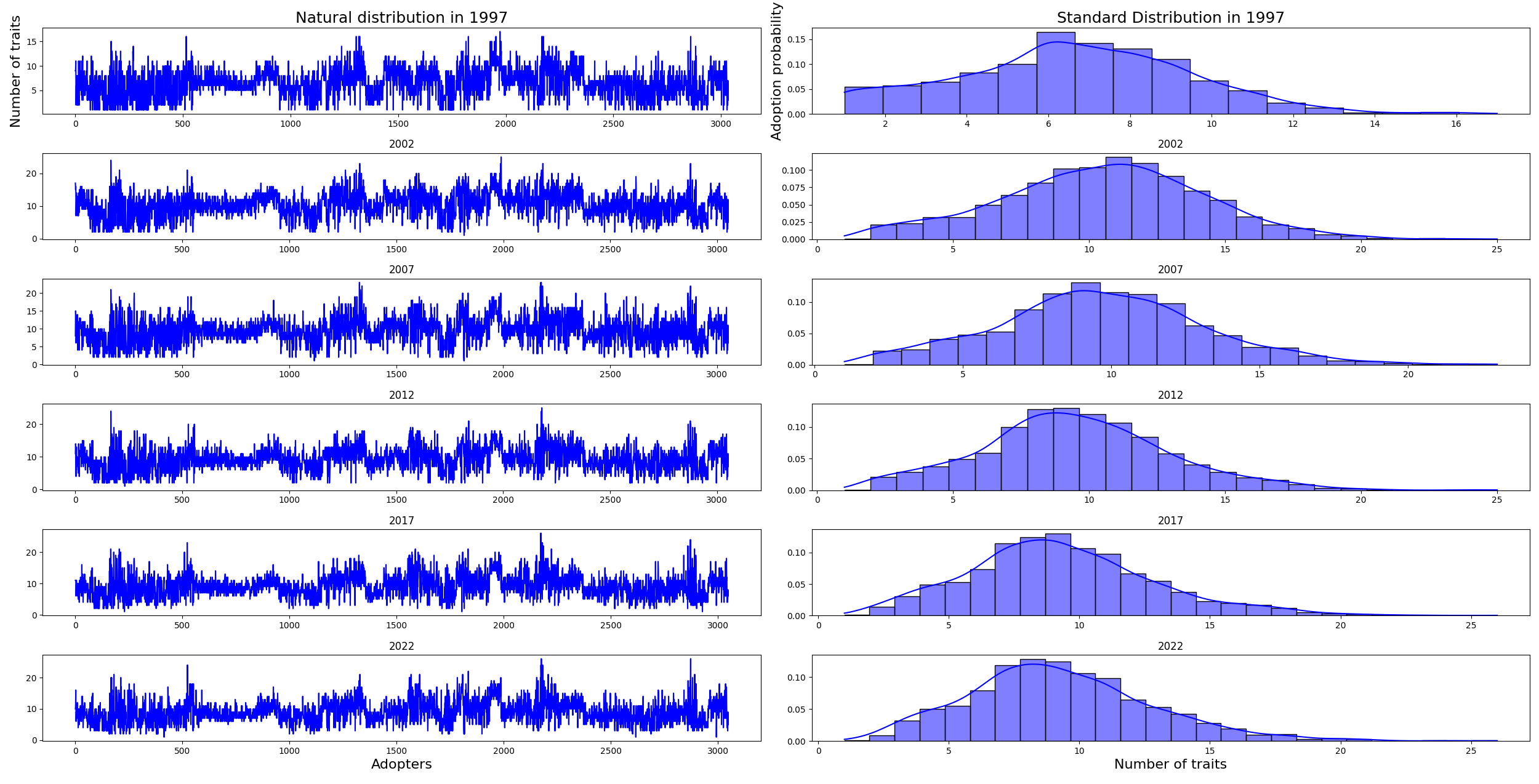}
  \caption{Adoption probability of traits by counties for all available data}
  \label{fig:adoptionden}
\end{figure}

 As it can be seen, our standard distributions show our data follows a normal\footnote{This distribution has been confirmed by using KS test for various distributions such as normal,lognormal, exponential, gamma and beta.} and the trend of adoption has been more towards trait\_8,trait\_9,trait\_10 and trait\_11.

\subsection*{Acknowledgments}
The author gratefully acknowledges his advisor, Professor Timothy M. Waring of The University of Maine, Orono, for carefully reviewing an earlier version of this manuscript and for his insightful comments and constructive suggestions.

\subsection*{Financial Support}
This project was supported by NSF (grant no. EPS-2019470).

\bibliography{Sources}

\begin{thebibliography}{10}

\bibitem{acerbi2015if}
Alberto Acerbi and Alex Mesoudi.
\newblock If we are all cultural darwinians what’s the fuss about? clarifying recent disagreements in the field of cultural evolution.
\newblock {\em Biology \& philosophy}, 30:481--503, 2015.

\bibitem{altman2019understanding}
Arie Altman and Alex Mesoudi.
\newblock Understanding agriculture within the frameworks of cumulative cultural evolution, gene-culture co-evolution, and cultural niche construction.
\newblock {\em Human Ecology}, 47:483--497, 2019.

\bibitem{bentley2004random}
R~Alexander Bentley, Matthew~W Hahn, and Stephen~J Shennan.
\newblock Random drift and culture change.
\newblock {\em Proceedings of the Royal Society of London. Series B: Biological Sciences}, 271(1547):1443--1450, 2004.

\bibitem{billiard2018stochasticity}
Sylvain Billiard and Alexandra Alvergne.
\newblock Stochasticity in cultural evolution: a revolution yet to happen.
\newblock {\em History and philosophy of the life sciences}, 40:1--24, 2018.

\bibitem{binford1963red}
Lewis~R Binford.
\newblock " red ocher" caches from the michigan area: A possible case of cultural drift.
\newblock {\em Southwestern Journal of Anthropology}, 19(1):89--108, 1963.

\bibitem{bowman2013introduction}
Gregory~R Bowman, Vijay~S Pande, and Frank No{\'e}.
\newblock {\em An introduction to Markov state models and their application to long timescale molecular simulation}, volume 797.
\newblock Springer Science \& Business Media, 2013.

\bibitem{boyd1988culture}
Robert Boyd and Peter~J Richerson.
\newblock {\em Culture and the evolutionary process}.
\newblock University of Chicago press, 1988.

\bibitem{boyd1996culture}
Robert Boyd, Peter~J Richerson, et~al.
\newblock Why culture is common, but cultural evolution is rare.
\newblock In {\em Proceedings-british academy}, volume~88, pages 77--94. Oxford University Press Inc., 1996.

\bibitem{cavalli1981cultural}
Luigi~Luca Cavalli-Sforza and Marcus~W Feldman.
\newblock {\em Cultural transmission and evolution: A quantitative approach}.
\newblock Number~16. Princeton University Press, 1981.

\bibitem{claidiere2014darwinian}
Nicolas Claidi{\`e}re, Thomas~C Scott-Phillips, and Dan Sperber.
\newblock How darwinian is cultural evolution?
\newblock {\em Philosophical Transactions of the Royal Society B: Biological Sciences}, 369(1642):20130368, 2014.

\bibitem{dunbar2007oxford}
Robin Ian~MacDonald Dunbar and Louise Barrett.
\newblock {\em Oxford handbook of evolutionary psychology}.
\newblock Oxford University Press, USA, 2007.

\bibitem{el2014cultural}
Claire El~Mouden, J-B Andr{\'e}, Olivier Morin, and Daniel Nettle.
\newblock Cultural transmission and the evolution of human behaviour: a general approach based on the price equation.
\newblock {\em Journal of evolutionary biology}, 27(2):231--241, 2014.

\bibitem{ellen20018}
Roy Ellen.
\newblock Cultural adaptation.
\newblock {\em The international Encyclopedia of anthropology}, 2018.

\bibitem{fog1999cultural}
Agner Fog.
\newblock {\em Cultural selection}.
\newblock Springer Science \& Business Media, 1999.

\bibitem{gulick1905evolution}
John~Thomas Gulick.
\newblock {\em Evolution, racial and habitudinal}.
\newblock Number~25. Carnegie institution of Washington, 1905.

\bibitem{henrich2001cultural}
Joseph Henrich.
\newblock Cultural transmission and the diffusion of innovations: Adoption dynamics indicate that biased cultural transmission is the predominate force in behavioral change.
\newblock {\em American anthropologist}, 103(4):992--1013, 2001.

\bibitem{henrich2016secret}
Joseph Henrich.
\newblock {\em The secret of our success: How culture is driving human evolution, domesticating our species, and making us smarter}.
\newblock princeton University press, 2016.

\bibitem{henrich2003evolution}
Joseph Henrich and Richard McElreath.
\newblock The evolution of cultural evolution.
\newblock {\em Evolutionary Anthropology: Issues, News, and Reviews: Issues, News, and Reviews}, 12(3):123--135, 2003.

\bibitem{hubbell2011unified}
Stephen~P Hubbell.
\newblock {\em The unified neutral theory of biodiversity and biogeography (MPB-32)}.
\newblock Princeton University Press, 2011.

\bibitem{kendal2018social}
Rachel~L Kendal, Neeltje~J Boogert, Luke Rendell, Kevin~N Laland, Mike Webster, and Patricia~L Jones.
\newblock Social learning strategies: Bridge-building between fields.
\newblock {\em Trends in cognitive sciences}, 22(7):651--665, 2018.

\bibitem{kimura1979neutral}
Motoo Kimura.
\newblock The neutral theory of molecular evolution.
\newblock {\em Scientific American}, 241(5):98--129, 1979.

\bibitem{lewens2015cultural}
Tim Lewens.
\newblock {\em Cultural evolution: conceptual challenges}.
\newblock OUP Oxford, 2015.

\bibitem{mesoudi2011cultural}
Alex Mesoudi.
\newblock {\em Cultural evolution: How Darwinian theory can explain human culture and synthesize the social sciences}.
\newblock University of Chicago Press, 2011.

\bibitem{mesoudi2021cultural}
Alex Mesoudi.
\newblock Cultural selection and biased transformation: two dynamics of cultural evolution.
\newblock {\em Philosophical Transactions of the Royal Society B}, 376(1828):20200053, 2021.

\bibitem{mesoudi2018cumulative}
Alex Mesoudi and Alex Thornton.
\newblock What is cumulative cultural evolution?
\newblock {\em Proceedings of the Royal Society B}, 285(1880):20180712, 2018.

\bibitem{richerson2008not}
Peter~J Richerson and Robert Boyd.
\newblock {\em Not by genes alone: How culture transformed human evolution}.
\newblock University of Chicago press, 2008.

\bibitem{smolla2021underappreciated}
Marco Smolla, Fredrik Jansson, Laurent Lehmann, Wybo Houkes, Franz~J Weissing, Peter Hammerstein, Sasha~RX Dall, Bram Kuijper, and Magnus Enquist.
\newblock Underappreciated features of cultural evolution.
\newblock {\em Philosophical Transactions of the Royal Society B}, 376(1828):20200259, 2021.

\bibitem{van2016primate}
Carel~P Van~Schaik.
\newblock {\em The primate origins of human nature}.
\newblock John Wiley \& Sons, 2016.

\end{thebibliography}

\bibliographystyle{plain}
\end{document}